\documentclass[11pt]{article}

\usepackage[utf8]{inputenc}
\usepackage[english]{babel}
\usepackage{amsmath}
\usepackage{subfiles} 
\usepackage{blindtext}
\usepackage{subfiles}
\usepackage{longtable}
\usepackage{multicol}
\usepackage{hyperref}
\usepackage{amsmath}
\usepackage{wrapfig}
\usepackage{lscape}
\usepackage[utf8]{inputenc}
\usepackage[affil-it]{authblk}
\usepackage{times}
\usepackage[T1]{fontenc}
\usepackage[plain]{fancyref}
\usepackage[toc,page]{appendix}
\usepackage{multicol}

\usepackage{graphicx}
\begin{document}

\title{Determining the Efficacy of Different Parameterization of the z-expansion}
\author[1]{E. Gustafson}
\author[1]{Y. Meurice}
\affil[1]{Department of Physics and Astronomy, The University of Iowa, Iowa City, IA, USA}	
\date{}
\maketitle

\abstract{We develop a methodology to test the accuracy of lattice extrapolations of the form factors in B decays only using experimental data. We test this methodology by comparing the BGL parameterization proposed by \cite{Boyd1996} and the BCL parameterization proposed by \cite{BLC} in the context of $B \rightarrow \pi \ell \nu$. Our results suggest that the BCL parameterization is a slightly better predictor of the low $q^2$ region than the BGL parameterization and that this methodology can be extended to other decays such as $B\rightarrow D\ell \nu$.}

\section{Introduction}
		It is important to have an accurate description of the form factors $f_+$ and $f_0$ in $B\rightarrow\pi \ell \nu$ because these form factors have a complicated dependence on the lepton transfer momentum ($q^2$). Furthermore lattice computations are limited in the energy regions that can be probed and therefore are limited in the $q^2$ region where values of the form factors can be computed. The first step to handle the $q^2$ dependence is using the conformal mapping variable:
\begin{equation}
z = \frac{\sqrt{t_+ - q^2} - \sqrt{t_+ - t_0}}{\sqrt{t_+ - q^2} + \sqrt{t_+ - t_0}}
\label{eq:zdefinition}
\end{equation}
first introduced by \cite{PhysRevD.3.2807,PhysRevD.4.725}. This transformation maps  the cut $q^2$ plane into the unit disk and is the foundation for the parameterizations of the form factors using the z-expansion. 

	There are two important parameterizations of the z-expansion. One parameterization proposed by \cite{Boyd1996}, 
\begin{equation}
f_+^{\text{BGL}}(q^2;t_0) = \frac{1}{B(q^2)\phi(q^2)} \sum_{i}^N a_n z(q^2;t_0)^n,
\label{eq:BGLparam}
\end{equation} uses the function $\phi(q^2)$ to impose a simple constraint on the coefficients $a_n$. A second parameterization proposed by \cite{BLC}, 
\begin{equation}
f_+^{\text{BCL}}(q^2;t_0) = \frac{1}{1 - q^2/m_{B^{\ast}}^2} \sum_n^{N-1} b_n (z^n - (-1)^{N-n}\frac{n}{N}z^N),
\label{eq:BCLparam}
\end{equation}
addresses the issues of the $(1/q^{2})^{1/2}$ falloff that is introduced by \fref{eq:BGLparam} and the threshold constraint imposed by conservation of angular momentum \cite{BLC}. 
\section{Methodology}
		Our method of testing different parameterizations is straightforward. The first step involves ensuring that the parameterizations satisfactorily describe the form factor over the full kinematic region. This is done by fitting the parameterization to all available experimental data related to the decay process. Once it is clear that the parameterizations can adequately describe the decay rate over the full kinematic region, several smaller regions of phase space are chosen, including a region where lattice calculations have determined values of the form factor. The decay rate is fitted to each expansion in these regions and then the fits are compared inside and outside of the fitted region. Comparisons inside the fitted region use the minimized reduced $\chi^2$ value while the comparisons outside the fitted region use the following quantity which we call the predictive measure:
\begin{equation}
X_p^2 = \frac{1}{N_{\text{unfitted}}} \sum_{i}^{N_{\text{unfitted}}} \frac{(\Delta B_i^{\text{exp}} - \Delta B_i^{\text{fit}})^2}{\sigma_i^{\Delta B}}.
\label{eq:XP2}
\end{equation}
$N_{\text{unfitted}}$ is the number of unfitted data points, $\Delta B_i^{\text{exp}}$ and $\Delta B_i^{\text{fit}}$ are, respectively, the experimental and fitted branching fractions corresponding to the $q_i^2$ to $q_{i+1}^2$ bin. In addition to comparing the accuracy of the different fits, it is also important to examine the stability of the coefficients in the parameterization. We examine the coefficients stability by using the best fit parameterization over the full kinematic range to generate a large number of bootstrap datasets.  The parameterizations are fitted to these "synthetic" data sets to establish how variable these coefficients will be to new measurements of the partial branching fractions.  

\section{Comparison of BGL and BCL parameterizations}
We examined the BGL and BCL parameterizations of the vector form factor $f_+(q^2)$ at  3 and 4 parameters (maximal order $z^2$ and $z^3$) for the BGL parameterization and 2 and 3 parameters (maximal order $z^2$ and $z^3$) for the BCL parameterization. These choices were made because they correspond to the minimal number of parameters and the next highest order that can adequately describe the form factors in the lattice regime; \cite{Bailey:2008wp} argue that a curvature term ($z^2$) is necessary to describe the lattice determinations of the form factors - this necessity is still be seen in \cite{Lattice:2015tia}. Although these parameterizations could be compared at equal number of parameters this choice biases the comparison because the curvature is different for the two parameterizations if the same number of parameters are used because the BCL expansion has an extra order in z that is included by the conservation of angular momentum \cite{BLC}. For example, the 2 parameter BCL expansion has the $z^2$ term included in the $b_1$ coefficient dependence:
\begin{equation}
f_+^{BCL} (q^2) = \frac{1}{1- q^2/m_{B^{\ast}}^2} (b_0 + b_1 ( z  + \frac{1}{2} z^2))
\end{equation}
while the 2 parameter BGL expansion, 
\begin{equation}
f_+^{BGL} (q^2) = \frac{1}{B(q^2) \phi(q^2)} (a_0 + a_1 z)
\end{equation}
does not possess a $z^2$ term. 
\subsection{Comparison of Lattice Region}

\begin{figure}[t]
\begin{multicols}{2}
\begin{center}
\includegraphics[width=0.35\textwidth]{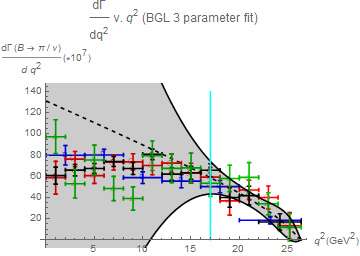}
\includegraphics[width=0.35\textwidth]{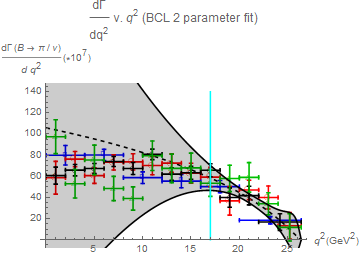}
\includegraphics[width=0.35\textwidth]{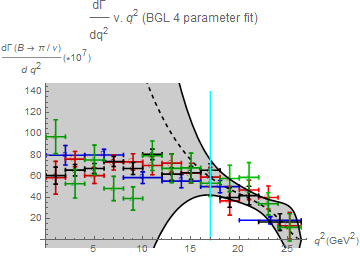}
\includegraphics[width=0.35\textwidth]{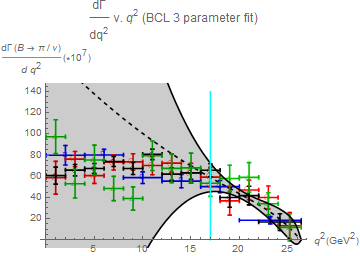}
\end{center}
\end{multicols}
\caption{fits to experimental data fitted $q^2 > 17 ~\text{GeV}^2$.  Index: (Blue squares: BaBar 2011 \cite{delAmoSanchez:2010af}, Red Circles: Belle 2011 \cite{Ha:2010rf}, Black diamond: BaBar 2012 \cite{Lees:2012vv}, Green Triangles: Belle 2013 results \cite{Sibidanov:2013rkk})}
\label{fig:expfit17264eo}
\end{figure}

\begin{figure}
\begin{multicols}{2}
\begin{center}
\includegraphics[width=0.35\textwidth]{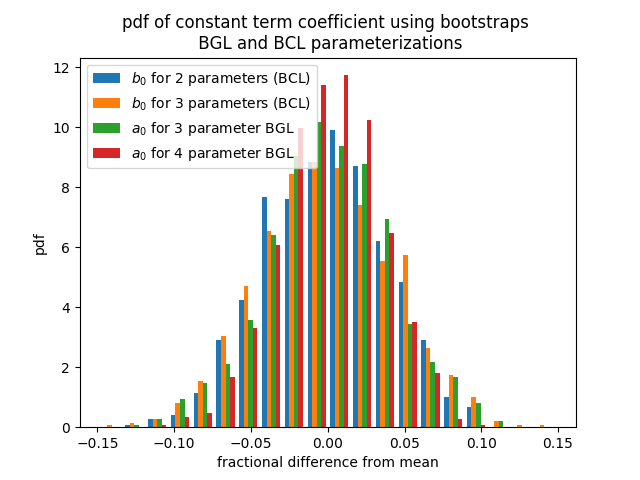}
\includegraphics[width=0.35\textwidth]{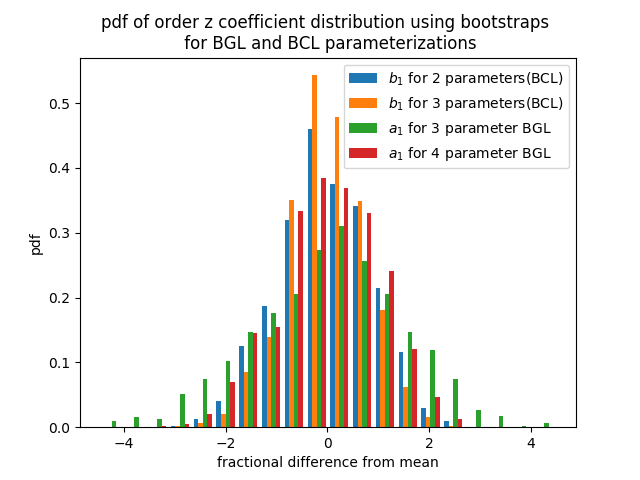}
\includegraphics[width=0.35\textwidth]{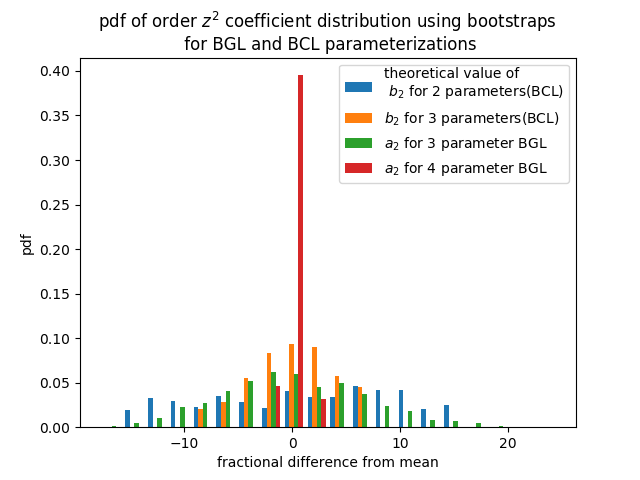}
\includegraphics[width=0.35\textwidth]{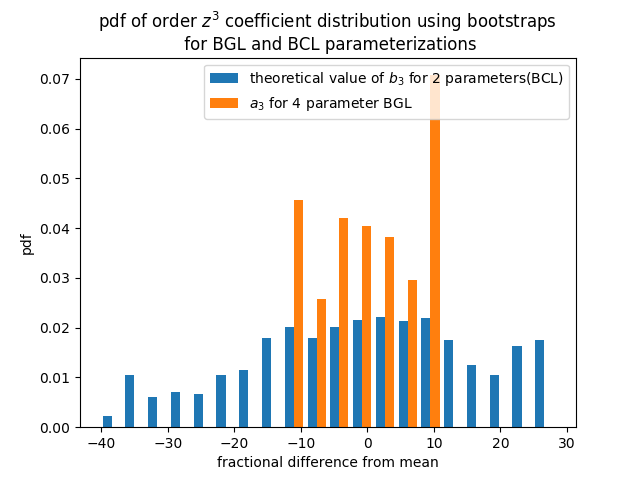}
\end{center}
\end{multicols}
\caption{results of fitting to lattice region with bootstrapped data sets. We used 1000 bootstrapped samples. Data outside of 20x mean value for the theoretical value of $b_2$ for the are not shown so that the resolution of the fitted coefficients can be seen}
\label{fig:bootstrap}
\end{figure}

The order $z^2$ fits corresponding to the lattice region is in the BGL case overestimates the partial branching fractions more than the BCL parameterization; but the order $z^3$ fit in both cases this over estimates the partial branching fractions outside of the lattice region (see \fref{fig:expfit17264eo}). The predictive measures $X_p^2$ are given in \fref{tab:Xp2valseo}. The significantly higher value for $X_p^2$ for the 4 parameter BGL fit and 3 parameter BCL fit compared to the 2 parameter BCL and 3 parameter BGL fits is likely due to overfitting. This can be seen in the bootstrap analysis; the constant, linear and quadratic terms for the 4 parameter BGL fit are gaussian but the cubic term is non normally distributed (see \fref{fig:bootstrap}),

\begin{table}
\begin{center}
\begin{tabular}{| c | c | c | c |}
\hline
order & type & 17 - 26.4 GeV$^2$ & 15 - 26.4 GeV$^2$\\ \hline
$z^2$ & BGL & 17.59 & 3.23 \\ \hline
$z^2$ & BCL & 7.97 &  2.62 \\ \hline
$z^3$ & BGL & 897  & 1.90 \\ \hline
$z^3$ & BCL & 48.5  & 4.33 \\ \hline
\end{tabular}
\end{center}
\caption{$X_p^2$ for a 3 and 4 parameter BGL and BCL expansions in the lattice and near lattice regions.}
\label{tab:Xp2valseo}
\end{table}
\subsection{Near Lattice Region}
A substantial difference in predictions occurs when the form factor parameterizations are fit to a region slightly larger (15 - 26.4 GeV$^2$) than the lattice region. The BCL and BGL parameterizations  all reasonably predict the low $q^2$ region. The predictive measure ($X_p^2$) for all four choices lie approximately between 2.0 and 4.5. This likely indicates that either overfitting is no longer as significant an issue because the data set has increased, or that this is an energy region where there is an interesting feature of the differential decay spectrum of $B \rightarrow \pi \ell \nu$ which occurs. 
\begin{figure}
\begin{multicols}{2}
\begin{center}
\includegraphics[width=0.35\textwidth]{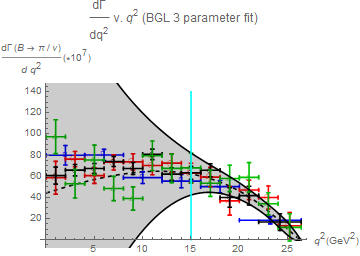}
\includegraphics[width=0.35\textwidth]{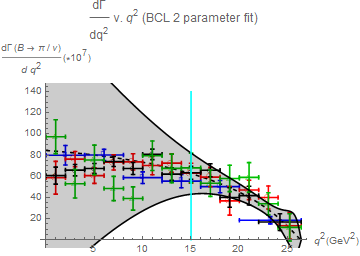}
\includegraphics[width=0.35\textwidth]{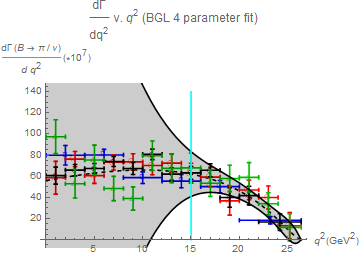}
\includegraphics[width=0.35\textwidth]{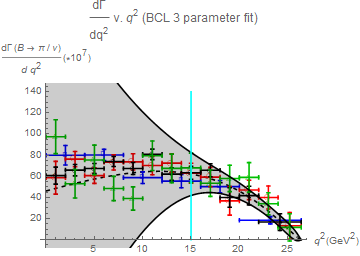}
\end{center}
\end{multicols}
\caption{fits to experimental data fitted $q^2 > 15 ~\text{GeV}^2$. Index: (Blue squares: BaBar 2011 \cite{delAmoSanchez:2010af}, Red Circles: Belle 2011 \cite{Ha:2010rf}, Black diamond: BaBar 2012 \cite{Lees:2012vv}, Green Triangles: Belle 2013 results \cite{Sibidanov:2013rkk})}
\label{fig:expfit15264eo}
\end{figure}

\section{Conclusions and Outlook}
We tentatively found that the BCL parameterization better predicts the low $q^2$ partial branching fractions than the BGL parameterization. The limiting factor in the efficacy of using the high $q^2$ partial branching fractions to predict the low $q^2$ partial branching fractions is the substantial statistical noise associated with the high $q^2$ partial branching fractions measured by experiment. This experimental noise (greater than 20-30 per-cent) is much larger than the uncertainties associated with the lattice determinations of the form factors (between 1 - 9 percent \cite{Lattice:2015tia}).  

There are two possible methods to address this issue. First would be to wait until LHCb or Belle II publishes their analysis of $B \rightarrow \pi \ell \nu$ that has better statistics. The second method would be to include the lattice results from \cite{Lattice:2015tia} to help constrain with $V_{ub}$ some of the parameters in the fit or if to use the lattice results to provide a priori estimates of the ratio of the fit coefficients in the z-expansion. The motivation for this comes from how the lattice and experimental results are determined. The partial branching fractions are binned; this results in a possible loss of some of the high $q^2$ behavior within each binned region. However the lattice results are not binned so more of the variability is visible. 

The substantial improvement of the predictions by slightly increasing the region to which we fit the data would suggest that a slight increase in the $q^2$ range for which we have lattice determinations of the form factors might provide a better idea on the shape of the form factor through the entire kinematic region. Therefore we suggest to find a method to extract a better signal for the high momentum modes.

It would be interesting to examine this process in other decay processes such as as $B \rightarrow D \ell \nu$ and $\Lambda_b \rightarrow \Lambda_{c} \ell \nu$. The each of these decays have analyses (\cite{Glattauer:2015teq} and \cite{Aaij:2017svr}) with available covariance matrices and partial branching fractions, although the volume of publicly data available is less than that of $B \rightarrow \pi \ell \nu$, this methodology can be used to help rule out other parameterizations of the form factors for these decays.
\section{Acknowledgements}
We would like to thank A. Schwartz for discussions regarding $B \rightarrow D$ decays. This research was supported in part by the Department of Energy under Award Number DOE DE-SC0010113
\bibliographystyle{ieeetr}
\bibliography{Tex/bibliography}

\end{document}